\documentclass[aps,pre,preprint]{revtex4-1}

\usepackage{graphicx}

\begin{document}
\title{Scale-Free Networks Hidden in Chaotic Dynamical Systems}
\author{Takashi Iba}
\email[]{iba@sfc.keio.ac.jp}
\homepage[]{http://web.sfc.keio.ac.jp/~iba/}

\affiliation{Faculty of Policy Management, Keio University, Fujisawa, Kanagawa 252-0882, Japan.}

\date{\today}

\begin{abstract}
In this paper,  we show our discovery that state-transition networks in several chaotic dynamical systems are ``scale-free networks,'' with a technique to understand a dynamical system as a whole, which we call the analysis for ``Discretized-State Transition'' (DST) networks; This scale-free nature is found universally in the logistic map, the sine map, the cubic map, the general symmetric map, the sine-circle map, the Gaussian map, and the delayed logistic map. Our findings prove that there is a hidden order in chaos, which has not detected yet. Furthermore, we anticipate that our study opens up a new way to a ``network analysis approach to dynamical systems'' for understanding complex phenomena.
\end{abstract}

% insert suggested PACS numbers in braces on next line

\maketitle

\section{Introduction}

Since time immemorial, human beings have sought to understand the relationship between ``order'' and ``chaos.'' Some of the earliest conceptions associated order with pattern and consistency, and chaos with formlessness and fluidity. Scientific inquiries over the past four decades, however, have revealed that a number of nonlinear systems often exhibit irregular and aperiodic behaviors; their sequence of the states never repeat even though they are governed by a simple, deterministic rule \cite{Poincare1890,Ulam1947}. This astonishing discovery of ``deterministic chaos'' blurs the line between traditional conceptions of pattern and constancy, and formlessness and fluidity.

On one hand, a number of phenomena that behave chaotically have been found empirically in various systems in the natural, technological, and social world \cite{Lorenz1963,Ruelle1971,May1976,Li1975,Mackey1977,Stutzer1980,Libchaber1982,Saperstein1984,Babloyantz1988,Olsen1990}. On the other hand, a number of mathematical models exhibiting chaos have been studied, and consequently some identifiable orders have been recognized inside chaos from a certain viewpoint. A chaotic attractor visualization appears complex but coherent, for example the visually beautiful trajectories in a ``phase space,'' which is a common representation for the evolution of a dynamical systems \cite{Strogatz1994,Devaney2003}. Besides, universal order has been found in bifurcation, which occurs when the control parameter of the system is changed \cite{Feigenbaum1978,Feigenbaum1980}. These discoveries of order within chaos, however, focus only on a partial trajectory, namely attractor, rather than the whole behavior of systems.

In this paper, we introduce a new technique that we call the analysis for ``Discretized-State Transition'' (DST) networks, in order to understand the whole behavior of dynamical systems rather than a partial behavior in an attractor. Then, with using the technique, we show that the state-transition networks in several chaotic dynamical systems are universally ``scale-free networks''\cite{Barabasi1999,Barabasi2003}, where there are quite many nodes having few links but also a few hubs that have an extraordinarily large number of links.

\section{Discretized-State-Transition (DST) Networks}

The technique for understanding the whole structure of behaviors in a dynamical system, the analysis for ``Discretized-State Transition'' (DST) networks, follows a two-step procedure: First, a continuous unit interval of variables is subdivided into a finite number of subintervals, ``states''; Second, a directed network is built with using these states, where a node represents a state and a link represents a transition from one state to a successive state. 

Concretely, first, we subdivide a unit interval of variables into a finite number of subintervals and call them ``discretized states,'' or simply ``states.'' The discretization is carried out by rounding the value of variables to $d$ decimal places. Accordingly, one obtains a finite set of states whose size is given by $\Delta=10^{-d}$. Mathematically, the map is described as a composite function as $x_{n+1} = h \big(f(x_{n}) \big) = h \circ f (x_{n})$, where the function $h(x)$ is for the rounding function, which is either the round-up, round-off, or round-down function.

 \begin{figure}
\includegraphics[width= 95mm]{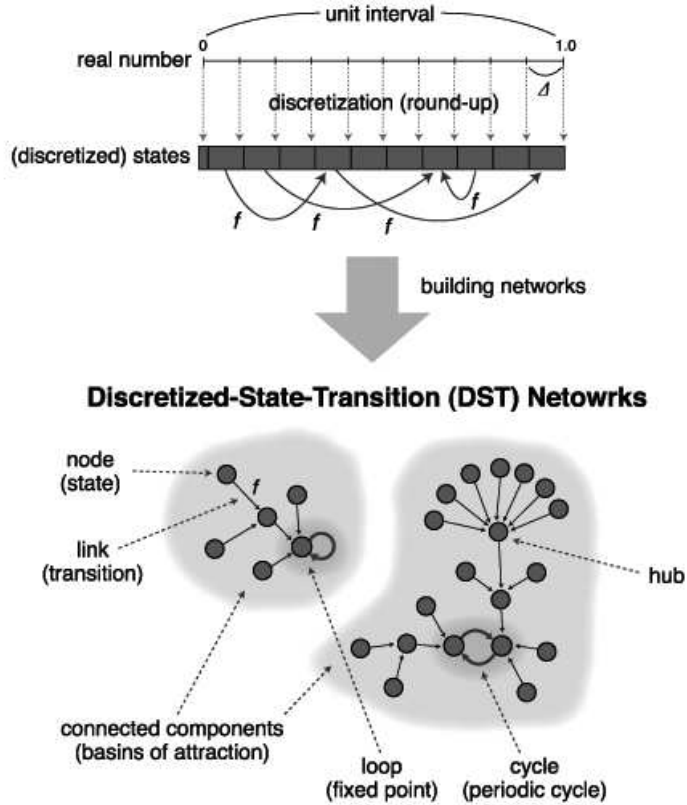}
\caption{Building Discretized-State Transition (DST) networks: To obtain the DST networks for a dynamical system, the continuous unit interval of variables is subdivided into a finite number of subintervals, which is called ``states,'' and then a directed network is build by connecting from one state to a successive state along the direction of the transition.}\label{Schematic}
\end{figure}

Second, we build a directed network by connecting a link from one state to a successive state determined by the dynamical system. The order of the network $N$, namely the number of nodes, is equal to $1/\Delta + 1$. The out-degree of every node must be always equal to 1 since it is a deterministic system, while the in-degree of each node can exhibit various values. The whole behavior is often mapped into more than one connected components, which represent ``basins of attraction.'' Each connected component must have only one loop or cycle, which represents ``fixed point'' or ``periodic cycle,'' where either is an ``attractor''. Fig.~\ref{Schematic} shows the simple schematic that summaries the technique, specifying the correspondent terms in the science of complexity with bracket under the terms in the network science.

\begin{figure}
\includegraphics[width= 70mm]{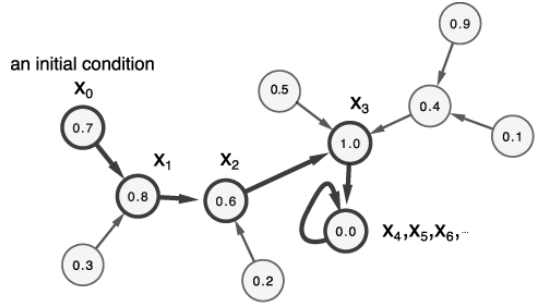}
\caption{DST networks and Numerical Simulation. Numerical simulation shows an instance of flow on the DST networks}\label{DifferenceFromSimulation}
\end{figure}

As a kind of directed network is occasionally called ``river network'', it may be helpful to understand a state-transition network with a metaphor of ``river.'' A node represents a geographical point in the river, and a link represents a connection from a point to another. The direction of a flow in the river is fixed. While there are no branches of the flow, however you can find confluences of two or more tributaries everywhere. To be exact, the network is not a usual river but ``dried-up'' river, where there are no flows on the riverbed. 

Determining a starting point and discharging water, you will see that the water flow downstream on the river network. That is a happening you witness when conducting a numerical simulation of system's evolution in time (Fig.~\ref{DifferenceFromSimulation}). Thus exploring a dynamical system with a numerical simulation means to observe an instance of flow starting from a point of the river, and then typical techniques of drawing a trajectory on phase space and bifurcation diagram are just to sketch out the instances. Our technique of a DST network, on the contrary, is to draw a ``map'' of the whole structure of the dried-up river networks, so one can overview the landscape from bird's-eye view. Thus the technique is much different from the typical technique of numerical simulations for understanding dynamical systems.

Even though there have been some pioneering studies of dynamical systems with finite-states \cite{Binder1986,Binder1998,Binder2003,Grebogi1988,Sauer1997} and studies of a state-transition network of discrete dynamical systems \cite{Wuensche1992,Kauffman1993,Kauffman1995,Wolfram1994,Wolfram2002,Peitgen2004}, little is known about the nature of the state-transition networks in chaotic dynamical systems.

\section{Scale-Free DST Networks in the Logistic Map}

\begin{figure}
\includegraphics[width= 85mm]{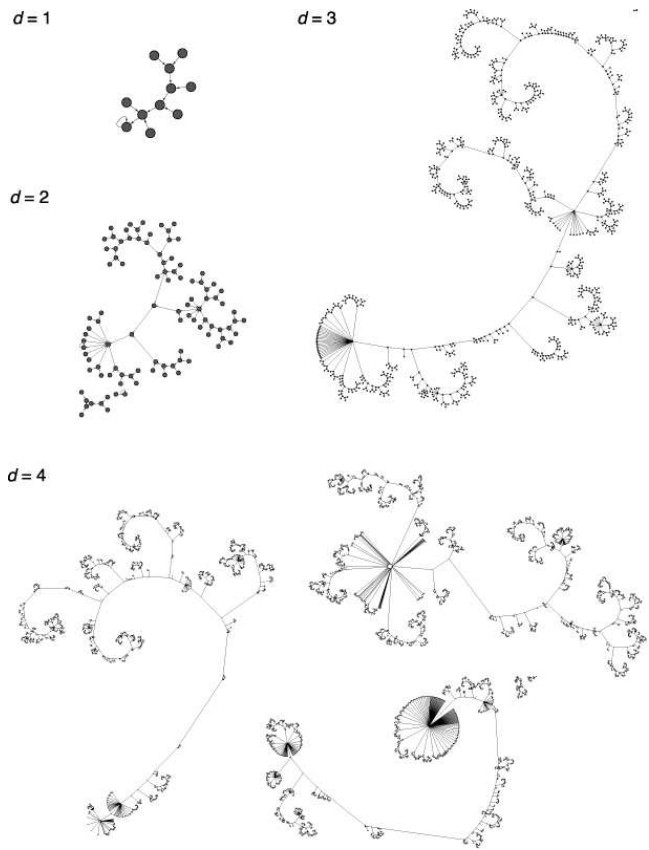}
\caption{The DST networks for the logistic map with $\mu=1$ in the case of round-up into $d$ decimal places ranging from $d=1$ to $d=4$. }\label{LogisticDST_d_net}
\end{figure}

\begin{figure}
\includegraphics[width= 90mm]{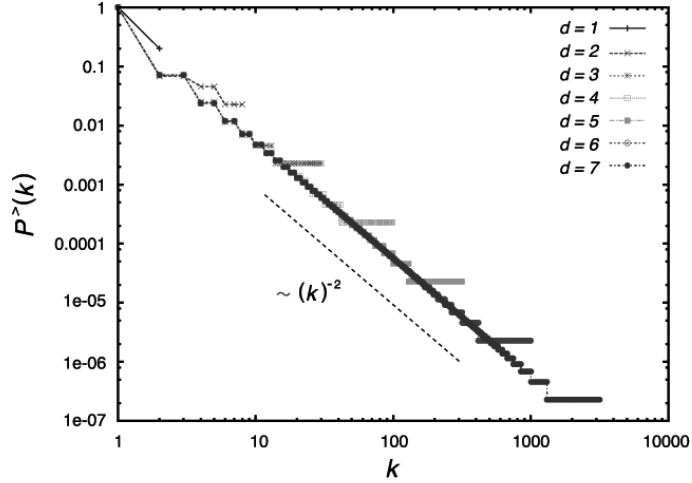}
\caption{The cumulative in-degree distributions of DST networks for the logistic map with $\mu=1$ in the case from $d=1$ to $d=7$. The dashed line has slope 2, implying that the networks are scale-free networks with the degree exponent $\gamma=1$, regardless of the value of $d$.}\label{LogisticDST_d_dist}
\end{figure}

We now move on to our discovery about chaotic dynamical systems with using the technique; that is, state-transition networks in several chaotic dynamical systems are ``scale-free networks.'' Such systems we found are mainly one-dimensional dynamical systems such as the logistic map, the sine map, the cubic maps, the general symmetric map, the Gaussian map, the sine-circle map, but also two-dimensional systems like the delayed logistic map.

To demonstrate our discovery, we start with a logistic map, which may be the simplest and most well known map. The function for the logistic map is given by $f(x)=4 \mu x ( 1 - x )$ , where $x$ is a variable between 0 and 1, and $\mu$  is a control parameter ranging from 0 to 1. Using this function, the value $x_{n+1}$ is determined by the previous value $x_{n}$ as $x_{n+1}= f (x_{n})$, and this is known as a simple population model with non-overlapping generations \cite{May1976}. In spite of the simple looking of the function, the model exhibits various kinds of behavior according to the control parameter $\mu$; for $0 \leq \mu < 0.75$ , all initial conditions attract to the fixed point; for $0.75 < \mu < 0.89...$,  $x$ oscillates as a periodic cycle; for $0.89...< \mu \leq 1.00$, the system exhibits chaos, namely the period is infinitely long.

Fig.~\ref{LogisticDST_d_net} shows DST networks of the logistic map with $\mu=1$ in the case of round-up into $d$ decimal places, ranging from $d=1$ to $d=4$, and fig.~\ref{LogisticDST_d_dist} shows the cumulative in-degree distribution for the states whose in-degree $k > 0$ , with $d$ ranging from 1 to 7. The dashed line in the figure has slope 2, therefore distributions follow a power law with the exponent $\gamma=1$ except in the regions where in-degree $k$ is extremely high. Consequently, the result implies that these networks are ``scale-free networks.''

\begin{figure}
\includegraphics[width= 90mm]{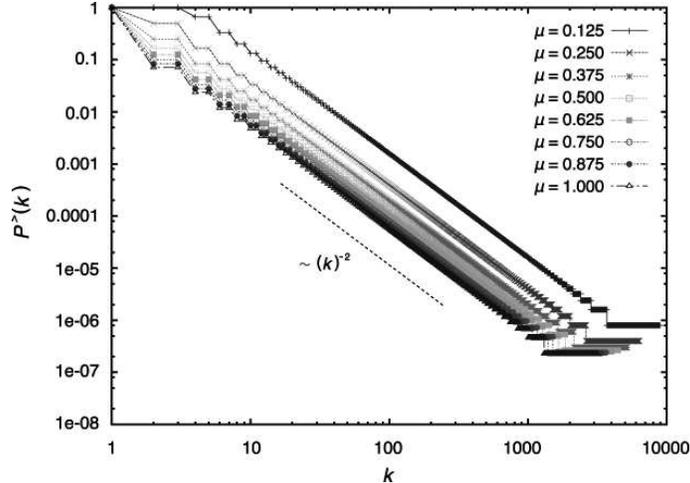}
\caption{The cumulative in-degree distributions of DST networks for the logistic map from $\mu=0.125$ to $1.000$ by $0.125$ in the case $d=7$. The dashed line has slope 2, implying that the networks are scale-free networks with the degree exponent $\gamma=1$, regardless of the value of the parameter $\mu$.}\label{LogisticDST_mu_dist}
\end{figure}

\begin{figure}
\includegraphics[width=85mm]{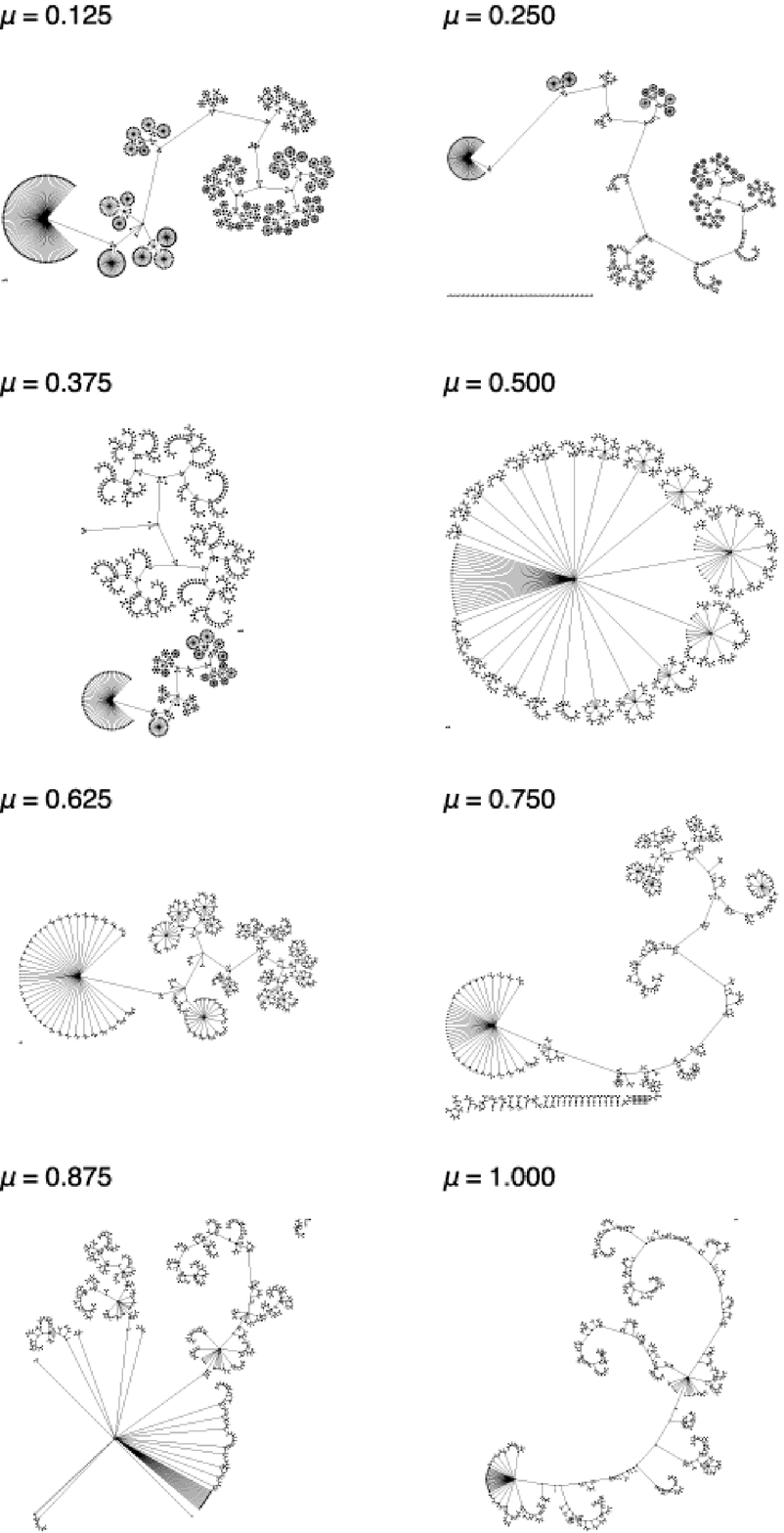}
\caption{Structural change of the DST networks for the logistic map in the case  $d=3$ with $\mu=0.125$, $\mu=0.250$, $\mu=0.375$, $\mu=0.500$, $\mu=0.625$, $\mu=0.750$, $\mu=0.875$, and $\mu=1.000$.}\label{LogisticDST_mu_net}
\end{figure}

Next, we investigated whether the scale-free nature of a DST network can be observed when varying the control parameter $\mu$, which governs the behavior of the system. Fig.~\ref{LogisticDST_mu_dist} shows the cumulative in-degree distribution for DST networks of the logistic map with $\mu$ ranging from 0 to 1 by $0.125$ in the case $d=7$. For all values, the distributions follow a power law with the exponent $\gamma=1$ except in the regions where in-degree $k$ is extremely high. The result means that the scale-free nature is independent of the parameter value. It also implies that the scale-free nature of state-transition networks is ascribable to the map function, not the chaotic regime of the system's behavior. Fig.~\ref{LogisticDST_mu_net} shows DST networks in the case $d=3$, which displays that the network structures are changed as the parameter $\mu$ is varied, even though the scale-free property remains.

Consequently, the results of our numerical explorations demonstrate that the DST networks of the logistic map are scale-free networks, independent of the control parameter $\mu$.

\section{Underlying Mechanism for Scale-Free DST Networks}

\begin{figure}
\includegraphics[width= 110mm]{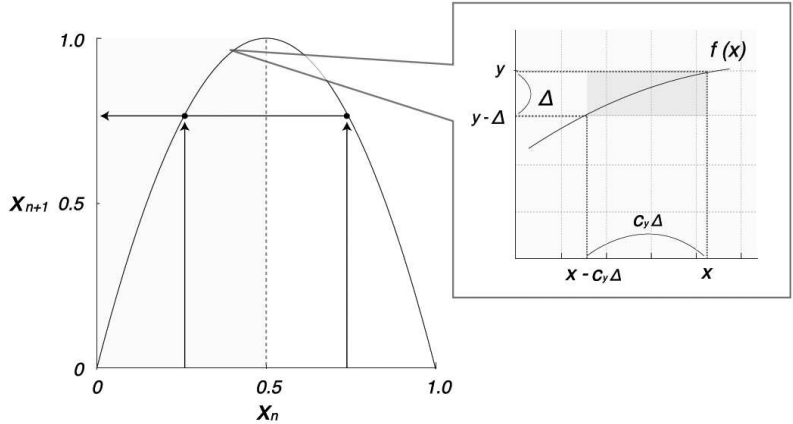}
\caption{The plot of the logistic map function with $\mu=1$. Note that the function is symmetric with respect to the critical point $x_{n}^{*}=0.5$, and then $x_{n+1}$ always yields two previous values of $x_{n}$ but the case $x_{n+1}=1$. The box shows relation between a subinterval on $y$-axis and its corresponding range on the $x$-axis for the logistic map function.}\label{Mathematical}
\end{figure}

We turn now to examine the underlying mechanism with mathematical consideration. Investigating the correlation between the value of $x$ and its in-degree $k_{x}$ by numerical computation, it turns out that in-degree $k_{x}$ becomes higher as the value of $x$ is getting to be larger. From the observation, it turns out that in-degree depends on the slope of the map function, and the hub states are located at the top of the parabola, whose slope is quite small. We shall describe a relation between $y$ and $x$ with using the size of subinterval $\Delta$ and the coefficient $c_{y}$, as shown in fig.~\ref{Mathematical}. Based on the observation that the in-degree becomes higher as $y$ is getting to be higher, the rank of state  $y$ is expressed as $r_{y}=(\mu-y)/\Delta + 1$. In addition, as pointed out in some literatures \cite{Newman2010,Sornette2004}, cumulative degree distribution can be calculated from the rank / frequency plots as $P^{>}(k)=r/N$, where $r$ is the rank and $N$ is a total number of nodes.

\begin{figure}
\vspace*{0.5cm}
\includegraphics[width= 85mm]{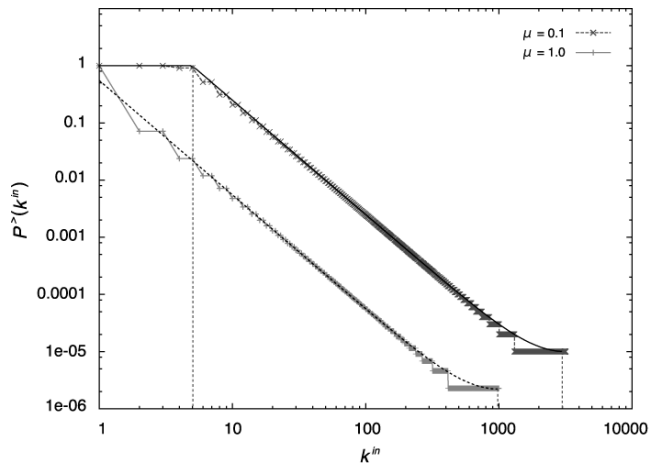}
\caption{Comparison between the results of numerical computations and mathematical predictions with $\mu=0.1$ and $\mu=1.0$ in the case $d=6$.}\label{SimulationMathematical}
\end{figure}

Replacing $r$ and $N$ of the cumulative distribution function $P^{>}(k)$, after some mathematical deduction (see more details in Appendix \ref{AppendixMath}), we obtain the equations for describing cumulative in-degree distribution of DST networks for the logistic map as follows:

\[ P^{>}(k) \propto \left(\frac{1}{k} + \mu \Delta k \right)^2,\]

\noindent where in-degree $k$ is in the range $0<k\leq k^{max}$ and $k^{max}=1/\sqrt{\mu  \Delta}$. Recall that  $0 < \mu \Delta < 1$, therefore the second term makes a large effect on the distribution only when $k$ is quite close to $k^{max}$. It means that the ``hub'' states will have higher in-degree than the typical scale-free network whose in-degree distribution follows a strict power law. Fig.~\ref{SimulationMathematical} shows the fitting between the results of numerical computation and its mathematical predictions. As a result, we come to a conclusion that the underlying mechanism presented above is valid for the model to explain how scale-free networks are formed.

Here, taking the limit of the equation, we have:

\[ \lim_{\Delta \rightarrow 0} P^{>}(k) \propto k^{-2}.\]

\noindent The result shows that the cumulative in-degree distribution becomes a power-law distribution in a strict sense, as the size of subintervals, $\Delta$, is getting to be infinitely small.

\section{Scale-Free DST Networks in Other Chaotic Dynamical Systems}

\begin{figure}
\includegraphics[width= 130mm]{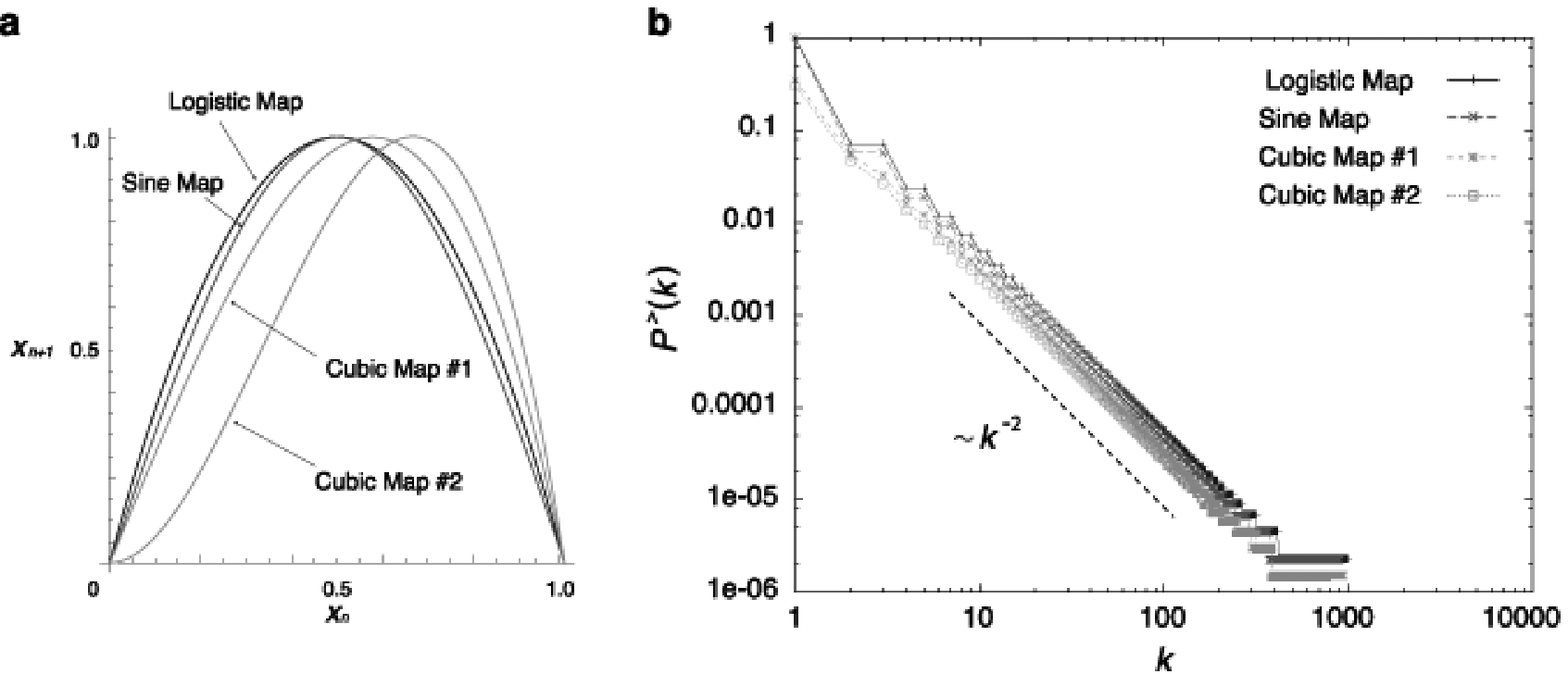}
\caption{(a) The plots for the logistic map, the sine map, and the cubic maps, $\sharp$1 and $\sharp$2; and (b) their cumulative in-degree distributions in the case $d=6$. The dashed line has slope 2, implying that the networks are scale-free networks with the degree exponent $\gamma=1$.}\label{SineCubicDist}
%\end{figure}

\vspace*{1cm}
%\begin{figure}
\includegraphics[width= 90mm]{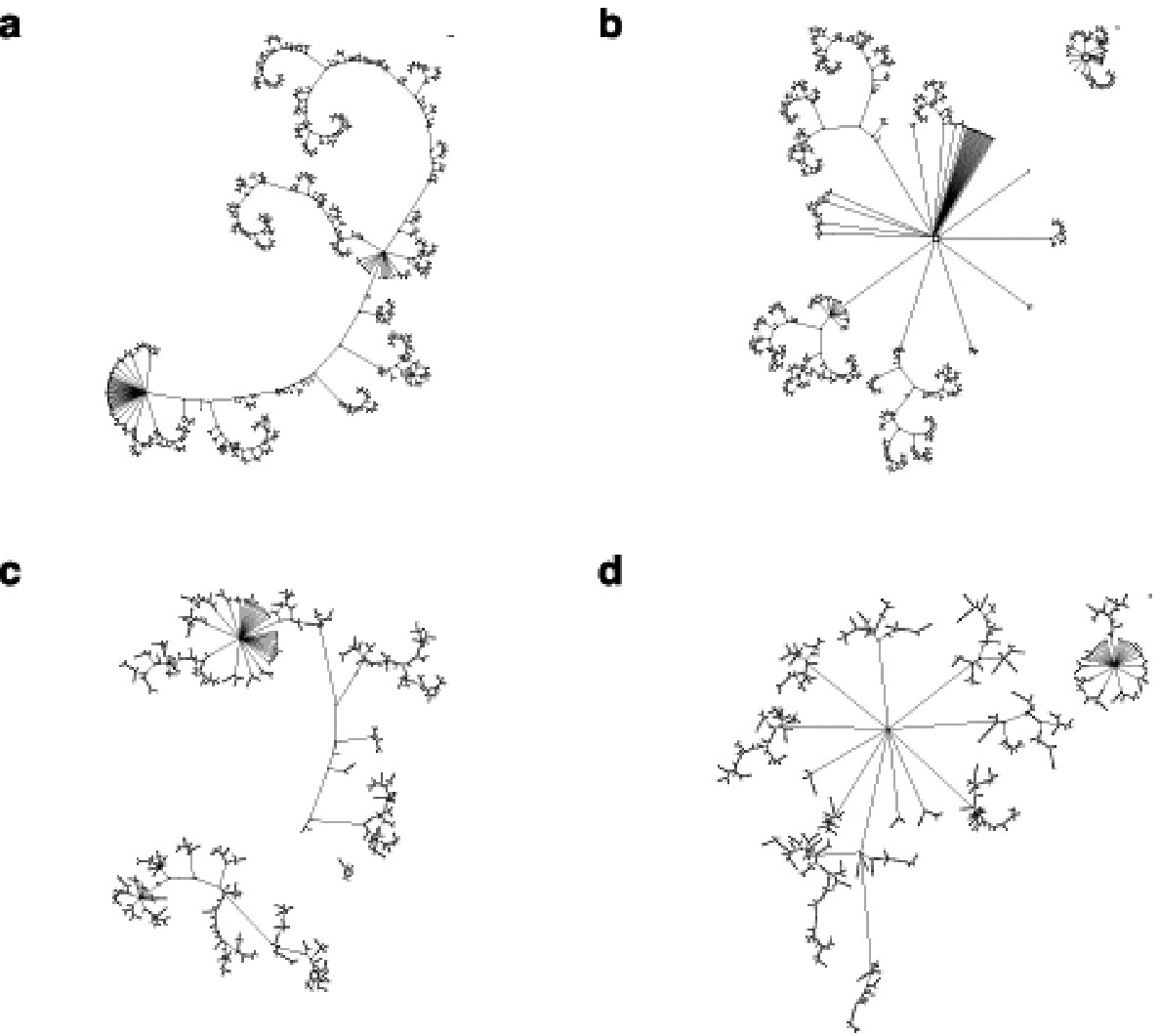}
\caption{Scale-free DST networks, in the case $d=3$, for (a) the logistic map with $\mu=1.0$; (b) the sine map with $\mu=1.0$ ; (c) the cubic map $\sharp$1 with $\mu=1.0$; and (d) the cubic map $\sharp$2 with $\mu=1.0$. }\label{SineCubicNet}
\end{figure}

Since we focused only on the logistic map so far, we now broaden the scope to other maps in order to show that the scale-free DST networks are universally found in several chaotic dynamical systems. Fig.~\ref{SineCubicDist} shows the plot of the following map functions and their cumulative in-degree distributions: the sine map $f(x)=\mu sin(\pi x)$, and cubic maps $f(x)=3 \sqrt{3} \mu x ( 1 - x^2 )/2$ and $f(x)=27 \mu x^2 (1-x)/4$. The result implies that the scale-free property of DST networks is not specific to the logistic map. Furthermore, the similarity in the degree distributions and its exponents among several systems suggests a new kind of ``universality'' of state transitions, which is different from the well-known universality of bifurcation \cite{Feigenbaum1978,Feigenbaum1980}. Examples of the DST networks for these maps are shown in fig.~\ref{SineCubicNet}.

\begin{figure}
\includegraphics[width= 130mm]{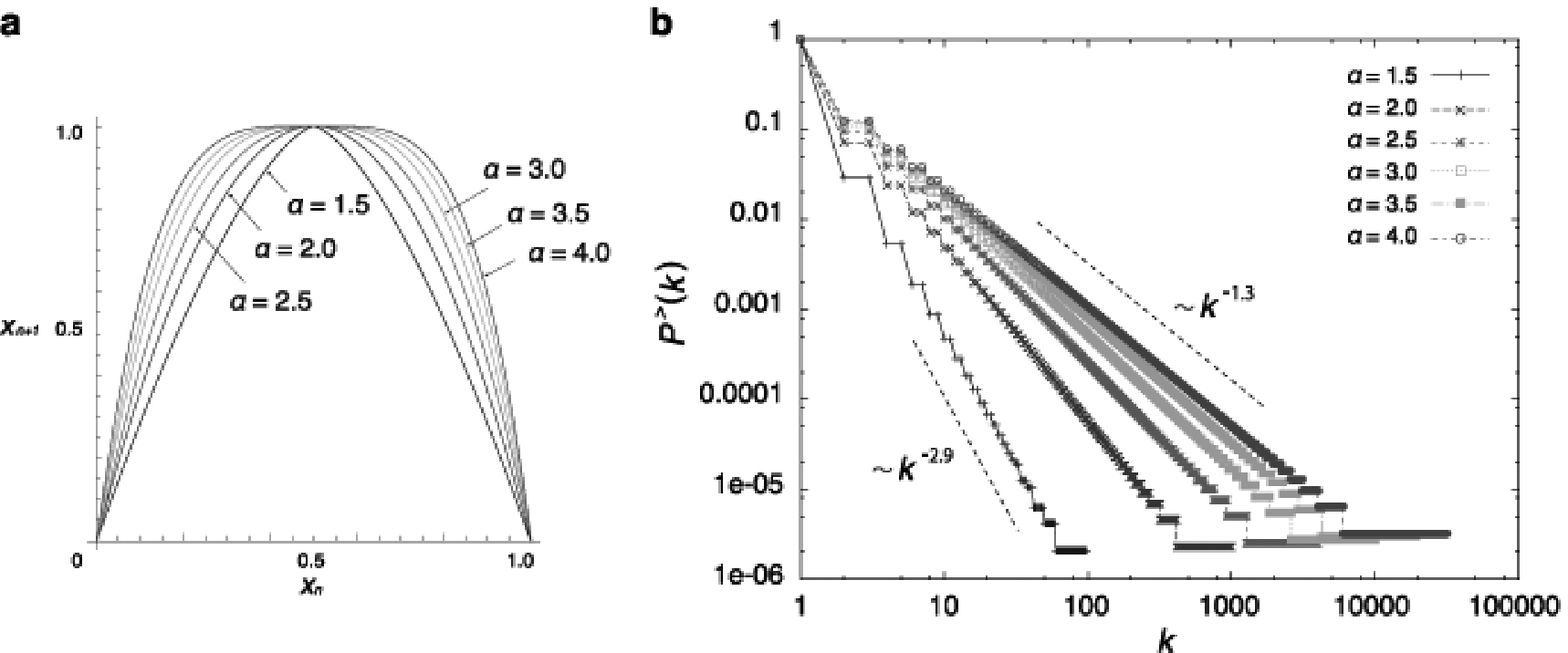}
\caption{(a) The plots for the general symmetric map with $\alpha$ ranging from $1.5$ to $4.0$ by $0.5$; and (b) their cumulative in-degree distributions in the case $d=6$. This figure shows that the parameter $\alpha$ influences the degree exponent $\gamma$, while the scale-free property is maintained.}\label{GeneralSymmetricDist}
%\end{figure}

%\begin{figure}
\vspace*{1cm}
\includegraphics[width= 90mm]{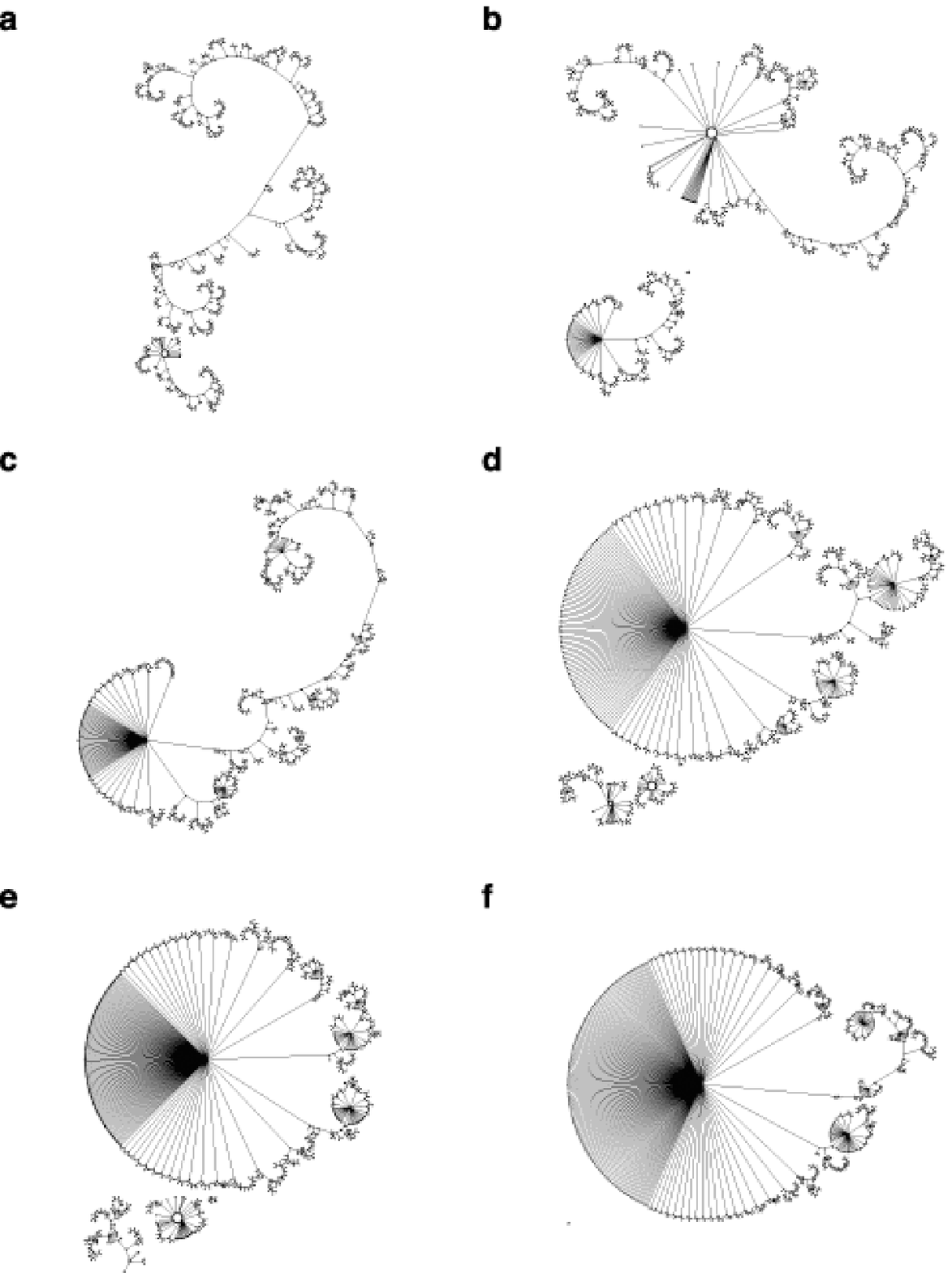}
\caption{Scale-free DST networks for the general symmetric map in the case $d=3$ with (a) $\alpha=1.5$; (b) $\alpha=2.0$; (c) $\alpha=2.5$; (d) $\alpha=3.0$; (e) $\alpha=3.5$; and (f) $\alpha=4.0$.}\label{GeneralSymmetricNet}
\end{figure}

Next, in order to understand how the flatness around the top of the parabola in a map function, we investigated the general symmetric map given by $f(x)=\mu \Big(1-|2x-1|^{\alpha} \Big)$. Fig.~\ref{GeneralSymmetricDist} 
shows the plot of the map function and their cumulative in-degree distributions, varying the value of $\alpha$ from 1.5 to 4.0 by 0.5. Note that the parameter $\mu$ is controlling the flatness around the top and the case $\mu=2$ is the same as the logistic map. The figure of the cumulative in-degree distribution shows that the parameter $\alpha$ influences the degree exponent $\gamma$; however the scale-free property is maintained in the all case of $\alpha \ge 1.5$. Examples of the DST networks for the map are shown in fig.~\ref{GeneralSymmetricNet}.

\begin{figure}
\includegraphics[width= 130mm]{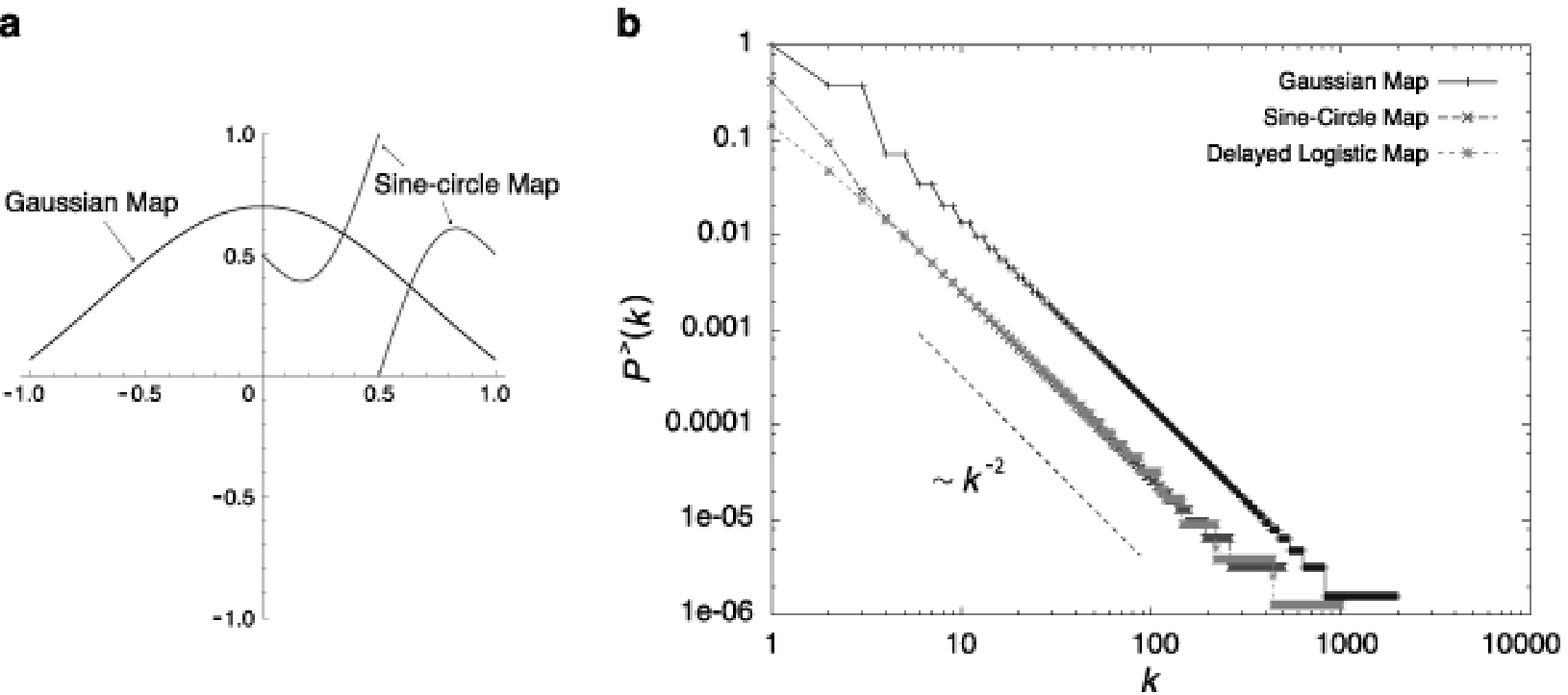}
\caption{(a) The plots for the Gaussian map and the sine-circle map; and (b) their cumulative in-degree distributions for the Gaussian map and the sine-circle map in the case  $d=6$ and the delayed logistic map in the case $d=3$. The dashed line has slope 2, implying that the networks are scale-free networks with the degree exponent $\gamma=1$.}\label{SineCircleGaussianDelayedDist}
%\end{figure}

\vspace*{1cm}
%\begin{figure}
\includegraphics[width= 90mm]{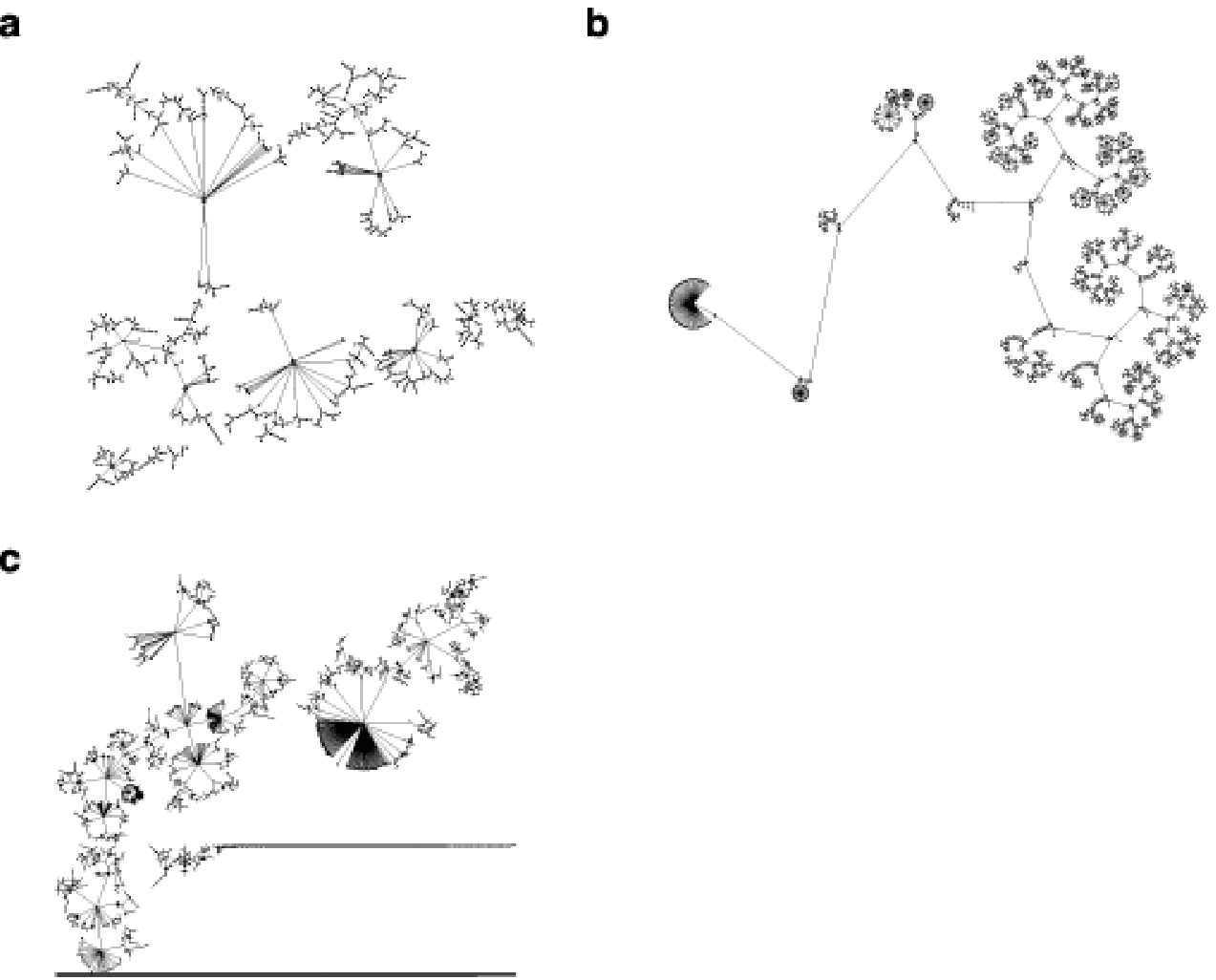}
\caption{Scale-free DST networks for (a) the sine-circle map with $a=4.0$ and $b=0.5$ in the case $d=3$; (b) the Gaussian map with $a=1.0$ and $b=-0.3$  in the case $d=3$; and (c) the delayed logistic map with $a=2.27$ in the case $d=2$.}\label{SineCircleGauusianDelayedNet}
\end{figure}

Finally, we explored the DST networks for the other types of map functions: exponential function, discontinuous function, and two-dimensional function (Fig.~\ref{SineCircleGaussianDelayedDist}). First, Gaussian map given by $f(x)=b + e^{-ax^{2}}$, which is an exponential function, is scale-free network in the certain range of the parameter $a$. Second, Sine-circle map \cite{Arnold1965} given by $f(x)=x + b - (a /2\pi) x ~~(mod ~1)$, which is discontinuous function, is also scale-free network. Third, the delayed logistic map \cite{Aronson1982} given by $f(x,y) = \Big( ax(1-y), x \Big)$, which is two-dimensional function, is scale-free network as well. In all cases, the in-degree exponent is universally $\gamma=1$. Examples of the DST networks for these maps are shown in fig.~\ref{SineCircleGauusianDelayedNet}.

%\vspace{0.5cm}

%\vspace{0.5cm}

\section{CONCLUSION}

Our discovery that state-transition networks of chaotic dynamical systems are organized in a certain way implies that our technique of mapping the transitions among discretized states is useful to capture the nature of the system as a whole. The recent development of the network science \cite{Newman2006,Caldarelli2007,Barabasi2009,Newman2010}  enables us to understand complex state-transition networks in light of a ``scale-free'' concept rather than random viewpoint. 

On the other hand, for network science, our study provides a demonstration of analyzing a network embedded in temporal behavior, where node is a ``state'' rather than either a substance or an actor. Note that, in such a network, more than one node cannot exist together at the same time. Network science has not studied this kind of temporally-embedded networks in much detail, and thus our study can be considered a step towards a new application of network analysis methods. 

Thus we anticipate that our study opens up a new way to a ``network-analysis approach to dynamical systems,'' where one can understand complex systems with methods and tools developed in the network science. 

\clearpage

\appendix
\section{Mathematical Proof of Scale-Free DST Networks for the Logistic Map}\label{AppendixMath}

As shown in the paper, our simulation results demonstrate that the DST networks of the logistic map are scale-free networks with any values of control parameter, $\mu$, and resolution for the discretization, $\Delta$. Here we provide a mathematical proof of our findings. For the purpose, we shall begin with calculating the number of the states whose in-degree is higher than zero, and then calculating the cumulative in-degree distribution.

\subsection{Number of the states whose in-degree is higher than zero}

Suppose that $f$ is the unimodal map of the interval $[0,1]$ that is concave downward, symmetrical with respect to the critical point, and $f(0)=f(1)=0$ (see Fig.~\ref{Mathematical}). While there are two solutions for the inverse function $f^{-1}$, we shall indicate a solution on the left hand of parabola as $f_{L}^{-1}$. Notice that, due to the symmetrical nature, in-degree $k$ is always an even number but the critical point.

In order to obtain the number of nodes who have incoming link, $N_{k>0}$, we shall describe a relation between $y$ and $x$ with using the size of subinterval $\Delta$ and the coefficient $c_{y}$  (see Fig.~\ref{Mathematical}). We then have $y=f(x)$ and $y - \Delta =f(x - c_{y} \Delta )$, and, with the inverse function, $x=f_{L}^{-1}(y)$ and $x - c_{y} \Delta = f_{L}^{-1} (y - \Delta)$. Cancelling $y$ from these equations gives the equation:
  
  \[ c_{y} = \frac{x - f_{L}^{-1} \big( f(x) - \Delta \big)}{\Delta}.\]

\noindent On the other hand, cancelling $x$ gives the equation:

  \[ c_{y} = \frac{f_{L}^{-1}(y) - f_{L}^{-1} ( y - \Delta)}{\Delta} .\]
  
\noindent Accordingly, since $k_{y}=2 c_{y}$  by definition, in-degree of the state $y$ is calculated as:
  
  \[ k_{y} = \frac{2}{\Delta} \left( f_{L}^{-1} - f_{L}^{-1} ( y - \Delta) \right) .\]
  
Let the coordinates of the point where $k=4$, namely $c_{y}=2$, be $(X,Y)$, and therefore it must satisfy that $Y=f(X)$ and $c_{Y}= \left(f_{L}^{-1} - f_{L}^{-1} ( Y - \Delta ) \right) / \Delta = 2$. We now consider the number of states for two ranges: $c_{y} \ge c_{Y}$ and $0 < c_{y} < c_{Y}$. First, the number of states in the range $c_{y} \ge c_{Y}$ is calculated as:

\[ N_{c_{y} \ge c_{Y}}  =  \left\{%
\begin{array}{ll}
\displaystyle \frac{\mu - Y}{\Delta} + 1 & \quad if ~~ Y \ge 0 \\[12pt]
\displaystyle \frac{\mu}{\Delta} + 1 & \quad  if ~~ Y < 0 .
\end{array}\right. \]

\noindent Second, the number of states in the range $0 < c_{y} < c_{Y}$ is calculated as:

\[ N_{0 < c_{y} < c_{Y}}  =  \left\{%
\begin{array}{ll}
\displaystyle \frac{X}{\Delta} + 1 &\quad   if ~~ X \ge 0 \\[12pt]
\displaystyle 0 & \quad if ~~ X < 0 .
\end{array}\right. \]

\noindent Note that, since the map function is concave downward, $f(0) = f(1) = 0$, and focusing the left hand of the parabola, there are only two combinations of the conditions: $X \ge 0 ~ \wedge ~ Y \ge 0$ and $X < 0 ~ \wedge~  Y<0$. Therefore, the number of states in the range $0 < c$, namely the number of the states whose in-degree is higher than zero, $N_{k>0}$, is given as follows:

\[ N_{k>0} =  \left\{%
\begin{array}{ll}
\displaystyle \frac{\mu + X - Y}{\Delta} + 2  & \quad if ~~ X \ge 0 ~ \wedge ~ Y \ge 0 \\[12pt]
\displaystyle \frac{\mu}{\Delta} + 1 & \quad if ~~ X < 0 ~ \wedge ~ Y < 0 .
\end{array}\right. \]

\subsection{Cumulative in-degree distribution}
As shown in some literatures \cite{Newman2010,Sornette2004}, cumulative degree distribution can be calculated from the rank / frequency plots as:

\[ P^{>} (k) = \frac{r}{N},\]

\noindent where $r$ is the rank and $N$ is a total number of nodes. Based on the observation that the in-degree becomes higher as $y$ is getting to be higher, the rank of state $y$ is expressed as:

\[ r_{y} = \frac{\mu - y}{\Delta} + 1.\]

Replacing $r$ and $N$ of the cumulative distribution function $P^{>}(k)$, we obtain:

\[ P^{>}(k_{y})  =  \left\{%
\begin{array}{ll}
\displaystyle \frac{\mu + \Delta - y}{\mu + 2 \Delta + X - Y}  & \quad  if ~~ X \ge 0 ~ \wedge ~ Y \ge 0 \\[12pt]
\displaystyle \frac{\mu + \Delta - y}{\mu + \Delta}  & \quad  if ~~ X < 0 ~ \wedge ~ Y < 0 .
\end{array}\right. \]

\subsection{Cumulative in-degree distribution in the case of the logistic map}

In the case of the logistic map $f(x)=4 \mu x (1-x)$, a solution of the inverse function is given by $f_{L}^{-1}(y) = \left( 1 - \sqrt{1- y/\mu} ~ \right) / 2$, and the coordinates of the point where $k=4$, $(X,Y)$, is given by: 

\[ (X,Y) = \left(\frac{1}{2} - \frac{| -1 + 16 \mu \Delta |}{16 \mu}, ~~\frac{\Delta}{2} - \frac{1}{64 \mu} + \mu - 4 \mu \Delta^{2} \right) .\]

Solving the equation of $k_{y}$ for $y$, we obtain:

\[ y = \mu - \frac{1}{4\mu k^{2}} - \frac{\Delta}{4} \left( -2 + \mu \Delta k^{2} \right)  .\]

Since $0 < \Delta \le 1/10$ is always satisfied in the condition of this paper, the condition $X \ge 0 ~\wedge~ Y \ge 0$ means $1/(8+16 \Delta ) \le \mu \le 1$; and the condition $X <0 ~\wedge ~Y < 0$ means $0 < \mu < 1 / (8+16\Delta)$. Thus, the number of states whose in-degree is higher than zero is given by:

\[ N_{c_{y}>0} =  \left\{%
\begin{array}{ll}
\displaystyle \frac{1 - 4 | -1 + 16 \mu \Delta | + 32 \mu (1 + \Delta + 8 \mu \Delta^{2} )}{64 \mu \Delta} & \displaystyle \quad if ~~ \frac{1}{8+16\Delta} \le \mu \le 1 \\[12pt]
\displaystyle \frac{\mu}{\Delta} & \displaystyle \quad if ~~ 0 < \mu < \frac{1}{8+16 \Delta} .\\
\end{array}\right. \]

\noindent Note that the maximum in-degree is observed at the highest point of the parabola $y^{*}=\mu$, namely at the critical point $x^{*}=1/2$, and the maximum in-degree is estimated by:

\[ k^{max} = \frac{1}{\sqrt{\mu \Delta}}. \]

Consequently, in the case $1/(8+16\Delta) \le \mu \le 1$, the cumulative in-degree distribution $P^{>}(k_{y})$ is given by:

\[ P^{>}(k) = \frac{16}{1 - 4 | -1 + 16 \mu \Delta | + 32 \mu ( 1+ 3 \Delta + 8 \mu \Delta^{2})} \left(\frac{1}{k} + \mu \Delta k \right)^{2} .\]

\noindent In the case $0 < \mu < 1/(8+16\Delta)$, the distribution is given by:

\[ P^{>}(k) =  \left\{%
\begin{array}{ll}
\displaystyle \frac{1}{4\mu (\mu + \Delta)} \left(\frac{1}{k} + \mu \Delta k \right)^{2}  & \displaystyle \quad if ~~ \frac{1-\sqrt{\Delta / \mu + 1}}{\Delta} < k\\[12pt]
\displaystyle 1 & \displaystyle \quad if ~~ 0 < k \le \frac{1-\sqrt{\Delta / \mu + 1}}{\Delta} .\\
\end{array}\right. \]

Thus, it is proved that the cumulative in-degree distribution of the DST network for the logistic map follows a law given by:

\[ P^{>}(k) \propto \left(\frac{1}{k} + \mu \Delta k \right)^{2}, \]

\noindent where in-degree $k$ is in the range $0<k\le k^{max}$ and $k^{max}=1/\sqrt{\mu \Delta}$. Recall that $0 < \mu \Delta < 1$, therefore the second term makes a large effect on the distribution, only when $k$ is quite close to $k^{max}$. It means that the ``hub'' states will have higher in-degree than the typical scale-free network whose in-degree distribution follows a strict power law. Fig.~\ref{SimulationMathematical} shows the fitness between the results of numerical computation and mathematical predictions.

In order to understand the distribution when the size of subintervals, $\Delta$, is enough small, we take the limit of $\Delta$ into 0. In the case $1/ (8+16 \Delta) \le \mu \le 1$, 

\[ \lim_{\Delta \rightarrow 0} P^{>}(k) = \frac{16}{-5 + 32 \mu + 128 \mu^{2}} k^{-2} ,\]

\noindent and in the case $0 < \mu < 1/(8+16\Delta)$,

\[ \lim_{\Delta \rightarrow 0} P^{>}(k) =  \left\{%
\begin{array}{ll}
\displaystyle \frac{1}{4 \mu^{2}} k^{-2} & \displaystyle \quad  if ~~ \frac{1}{2\mu} < k \\[12pt]
\displaystyle 1 & \displaystyle \quad if ~~ 0 < k \le \frac{1}{2\mu} .
\end{array}\right. \]

\noindent Thus, it is clear that the cumulative in-degree distribution of the DST network for the logistic map follows a strict power law when $\Delta$ is enough small as follows:

\[ P^{>}(k) \propto k^{-2}. \]

\noindent In conclusion, the DST networks for the logistic map become ``perfect'' scale-free networks whose degree distribution strictly follows a power law with the exponent $\gamma=1$, as the size of subintervals is getting smaller.

\begin{acknowledgments}
I would like to thank to K. Shimonishi for collaborating on new ways of studying chaos. I also want to thank to B. Waber and C. Rolbin for helping to improve the manuscript, S. Matsukawa, K. Nemoto, N. Masuda, and Y. Shikano for helpful discussion, and T. Malone and P. Gloor for providing a research environment for@network analysis at MIT Center for Collective Intelligence.
\end{acknowledgments}

\end{document}